\documentclass[10pt]{IEEETran}

\def \be {\begin{equation}}
\def \ee {\end{equation}}

\def \nn {\nonumber}

\usepackage{ifpdf}

\usepackage[dvips]{graphicx}
\usepackage{color}

\usepackage[cmex10]{amsmath}
\usepackage{amssymb}

\begin{document}

\title{Task-Space Consensus of Networked Robotic Systems: Separation and Manipulability}

\author{Hanlei~Wang and Yongchun~Xie %\emph{IEEE Member}
        % <-this % stops a space
%\thanks{This work was supported by the National Natural Science Foundation of China under Grant 61374060, the National Key Basic Research
% Program (973) of China under Grant 2013CB733100, and the China Scholarship Council.}
\thanks{The authors are with the Science and Technology on Space Intelligent Control Laboratory,
Beijing Institute of Control Engineering,
Beijing 100094, China (e-mail: hlwang.bice@gmail.com; xieyongchun@vip.sina.com).% <-this % stops a space
}
}

\maketitle
%\IEEEpeerreviewmaketitle

\begin{abstract}
%\boldmath
In this paper, we investigate the task-space consensus problem for multiple robotic systems with both the uncertain kinematics and dynamics and address two main issues, i.e., the separation of the kinematic and dynamic loops in the case of no task-space velocity measurement and the quantification of the manipulability of the system. We propose an observer-based adaptive controller to achieve the manipulable consensus without relying on the measurement of task-space velocities, and also {formalize the concept of manipulability} to quantify the degree of adjustability of the consensus value. The proposed adaptive controller employs a new distributed observer that does not rely on the joint velocity and a new kinematic parameter adaptation law with a distributed adaptive kinematic regressor matrix that is driven by both the observation and consensus errors. In addition, it is shown that the proposed controller has the separation property, which yields an adaptive kinematic controller that is applicable to most industrial/commercial robots. The performance of the proposed observer-based adaptive schemes is shown by numerical simulations.
\end{abstract}


\section{Introduction}

Networked robotic systems have many potential applications such as cooperative manipulation, planet/field exploration, and teleoperation. This motivates the active research on the control of networked robotic systems in recent years (see, e.g., \cite{Rodriguez-Angeles2004_TCST,Chopra2006,
Cheng2008_ICSMC,Cheah2009_Aut,Nuno2011_TAC,
Mei2012_Aut,Wang2013a_Aut,Wang2013b_Aut,Wang2013_TAC,
Liu2012_TRO,Liu2013_AUT,Abdessameud2014_TAC,Wang2014_TAC,Ghapani2014_ACC,Liu2015_JFI}).
A fundamental control problem for networked robotic systems that is actively studied is consensus, in which case all systems are expected to reach agreement concerning certain variables. The major difficulty involved, as is frequently mentioned in the literature, is the nonlinearity and uncertainty of the system model.

The consensus schemes for networked robotic systems can generally be grouped, in accordance with the interaction graphs among the robotic systems, into two categories. The first category of schemes (e.g., \cite{Chopra2006_TRO,Lee2006_TRO,Cheng2008_ICSMC,Ren2009_IJC,Liu2013_AUT,Aldana2015_IJRNC}) achieves the consensus of robotic systems on undirected interaction graphs. In the case that there are gravitational torques in the system, the control schemes in \cite{Chopra2006_TRO,Lee2006_TRO,Ren2009_IJC,Aldana2015_IJRNC} require the exact knowledge of the gravitational torques to ensure the consensus. This dependence on knowing the gravitational torques is removed in \cite{Cheng2008_ICSMC,Liu2013_AUT} thanks to the employment of adaptive schemes. The second category of schemes (e.g., \cite{Chopra2006,Nuno2011_TAC,Mei2012_Aut,Wang2013_TAC,Wang2014_TAC,Abdessameud2014_TAC}) achieves the consensus of the robotic systems on the more general directed graphs. As is described/shown in \cite{Nuno2010_Aut,Nuno2011_TAC}, the adaptive version of the scheme in \cite{Chopra2006} gives rise to the outcome that all systems' positions converge to the origin in the presence of gravitational torques. This deficiency has been conquered by the adaptive scheme in \cite{Nuno2011_TAC}, and other relevant results appear in \cite{Min2012_SCL,Mei2012_Aut}. The adaptive scheme in \cite{Wang2014_TAC}, by employing the integral-sliding control action, achieves the (stability guaranteed) consensus of the systems with the final consensus value being explicitly expressed in terms of the initial systems' positions, and in fact, this adaptive scheme realizes the scaled weighted average consensus of the systems. The case of time-varying communication delays is considered in \cite{Abdessameud2014_TAC} where a small-gain-based consensus scheme is proposed. However, all the results above only take into account the dynamic uncertainties.

When the robotic system performs tasks given in the Cartesian space, kinematic uncertainties (e.g., the lengths of the robot links may not be accurately known) possibly occur \cite{Cheah2003_TRA,Cheah2006_IJRR,Cheah2006_TAC}. Therefore, various adaptive control algorithms are proposed to accommodate the kinematic uncertainties, using the estimated Jacobian matrix \cite{Cheah2006_IJRR,Cheah2006_TAC}. The consensus schemes with consideration of the uncertain kinematics or both the uncertain kinematics and dynamics appear in \cite{Cheng2008_MSC,Liu2010_IROS,LiuX2014_MECH,Liu2015_TMECH} (with the interaction graph being undirected) and in \cite{Wang2013_TAC,Liu2015_JFI} (with the interaction graph being directed and strongly connected). Motivated by the well-recognized fact that the task-space velocity measurement usually involves too much noise (due to the noisy nature of the task-space position measurement), the work in \cite{WangL2014_IJC} gives an observer-based adaptive consensus scheme that does not require the task-space velocity measurement, where the observer explicitly relies on the joint velocity. The second consensus scheme in \cite{Liu2015_TMECH} (which extends \cite{Cheah2006_IJRR} to the case of teleoperator systems) avoids the task-space velocity measurement at the expense of overparametrization and under the assumption that the communication delay is absent. %However, in this observer-based scheme, the consensus error information is not reflected in the kinematic parameter adaptation, and in addition the communication delays are not considered.
{While these results are effective in the case of an open torque design interface, in practice, however, most industrial/commertial robots do not provide this design interface and typically only the joint velocity (or position) command can be designed (see, e.g., \cite{Grotjahn2002_IJRR})}. %{Specifically, consider a group of robotic systems with uncertain kinematics and dynamics and with an unmodifiable joint servoing controller [typically PI (proportional-integral) velocity or PID (proportional-integral-derivative) position controller] in terms of the joint-space tracking error, and suppose that the joint servoing process is fast enough.
One solution to this problem in the context of a single robotic system with the availability of task-space velocity measurement is the separation approach in \cite{Wang2017_TAC}, yet no attempts have been devoted to extending this approach to address the task-space consensus problem for multiple robotic systems without involving task-space velocity measurement.

%Then we may wonder how to design a distributed adaptive kinematic consensus controller without involving task-space and joint-space velocity measurement and without directly involving task-space position measurement so that the task-space consensus with enhanced robustness can be ensured. The above mentioned approaches cannot resolve this problem appropriately due to the dependence on the modification of the low-level feedback controller structure, and the joint velocity measurement of the kinematic parameter adaptation law.

Another important issue concerning networked robotic systems with an external stimuli or human input action is what is referred to as manipulability %\footnote{``Dynamic manipulability'' also appears in many other contexts with strikingly different emphasis (e.g., measuring the performance by the relation between the end-effector acceleration and the joint torque of robot manipulators \cite{Yoshikawa1985_ICRA} and intuitively describing the properties of a virtual object in the field of computer engineering where the virtual object is often referred to as ``dynamically manipulable'' or ``interactively manipulable''); the concept here is associated with an interactive input (for instance exerted by a human operator) and an output that are related by a dynamic network of robotic systems with ``dynamic'' being contrast with ``static''.}
 (which can be intuitively interpreted as the degree of the adjustability of the consensus equilibrium), and this issue has not been {formally/systematically studied (has been implicitly used with no rigorous justification though---see, e.g., \cite{Aldana2015_IJRNC,Nuno2017_IJACSP})} in the previous work, especially in the presence of system uncertainties. The manipulability is particularly significant in a teleoperator system (see, e.g., \cite{Hokayem2006_AUT}) in that it is tightly related to the operator's physical feeling and perception concerning the teleoperator. The well-accepted concept of transparency \cite{Lawrence1993_TRA} is known to be a measure towards the degree of ``telepresence'' and it is concerning the achieved impedance relation in frequency domain between the applied force and velocity. This frequency domain description is, however, difficult to be extended to consider the nonlinearity of the dynamics of the teleoperator and also lacks the rigorous justification of why and how the human operator can adjust the equilibrium point of the teleoperator with {large or small amount of control efforts} (i.e., corresponding to much or less fatigue of the human operator).

In this paper, we propose a new observer-based adaptive scheme with the property of separation for achieving the task-space consensus of the networked robotic systems with both the uncertain kinematics and dynamics and with constant communication delays, and {formalize the concept of manipulability [for rigorously addressing which kind of controllers enables the human operator to adjust the consensus equilibrium with large (or small) amount of control efforts]} to quantify the degree of adjustability of the consensus value. Both the issue of designing adaptive kinematic schemes and that of the manipulability of the system are explicitly addressed. Specifically, the proposed new scheme employs 1) a new task-space observer that relies on the joint reference velocity rather than joint velocity, in contrast to the joint-velocity-dependent observer in \cite{WangL2014_IJC}, 2) the inverse Jacobian feedback control inspired by the results for a single robotic system \cite{Wang2017_TAC,Ma1995_ICRA},  %and ({cite Nuno and Wang's work}),
unlike most existing task-space consensus schemes (e.g., \cite{Wang2013_TAC,WangL2014_IJC,LiuX2014_MECH,Liu2015_TMECH}), and 3) a new kinematic parameter adaptation law with a distributed adaptive kinematic regressor, which is driven by both the observation error and consensus error (in contrast with \cite{WangL2014_IJC} where the kinematic parameter adaptation law is driven by the observation error only). These features
 result in the separation of the kinematic and dynamic loops while the existing consensus schemes (e.g., \cite{WangL2014_IJC,Liu2015_TMECH}) do not enjoy this property due to the coupling between the kinematic and dynamic loops.

 The main benefit of the new task-space observer without involving the joint velocity measurement as well as the inverse Jacobian feedback design is enhanced robustness with respect to the measurement noise and guaranteed separation (rather than the complete avoidance of joint velocity measurement in the torque control input).
  Enhanced robustness lies in the avoidance of joint velocity measurement in both the kinematic parameter adaptation law and the observed quantity of the task-space velocity since this significantly reduces the degree of involving joint velocity measurement (not completely avoided though) in the designed control torque. The separation property leads us to derive an adaptive kinematic controller applicable to the case of multiple robotic systems with an unmodifiable joint servoing controller (e.g., most industrial/commercial robots). The separation property achieved as well as the separation stability analysis can be considered as an extension of that for a single robotic system in \cite{Wang2017_TAC} to the case of multiple robotic systems without task-space velocity measurement, and this extension is realized by designing a new distributed task-space observer and using a new distributed adaptive kinematic regressor. In addition, our control scheme avoids the overparametrization problem and can conveniently handle the communication delays, in contrast with the second scheme in \cite{Liu2015_TMECH}. Another work given in \cite{Aldana2014_JFI} presents a task-space consensus controller that requires neither the task-space nor joint-space velocity measurement, but this is achieved by requiring that the interaction graph is undirected and the system model is exactly known while our result considers directed graphs and does not rely on the exact knowledge of the system model. %We also discuss the issues associated with the application of the proposed scheme to bilateral teleoperation.

%{In summary, the major contribution of this work is that by employing a new distributed adaptive kinematic regressor and a new distributed task-space observer which is coupled to the robotic system, we develop a distributed adaptive scheme that can achieve the task-space manipulable consensus (see the discussion in Remark 3) and separation of the kinematic and dynamic loops of networked robotic systems without involving task-space velocity measurement, facilitating the application to networked robotic systems with an unmodifiable joint servoing controller yet admitting the design of the joint velocity command (e.g., most industrial/commercial robots).}

In summary, {the main contribution of our study here, as compared with the existing results for networked Euler-Lagrange systems or robotic systems (e.g., \cite{Nuno2011_TAC,Wang2014_TAC,Wang2013_TAC,WangL2014_IJC,Liu2015_TMECH}) and also those in the context of bilateral teleoperation (see, e.g., \cite{Lee2006_TRO,Nuno2009_IJRR}), is to {formalize the concept of manipulability} followed by the systematic manipulability analysis concerning networked robotic systems and to address the case of no task-space velocity measurement by developing a new task-space observer. In particular,
 \begin{enumerate}
 \item we {rigorously show that the gain of the integral action concerning the sliding vector (i.e., the weighted sum of the velocity and neighbor-to-neighbor position consensus errors) acts as a qualified measure of manipulability of the closed-loop system}, and this provides a tuning freedom concerning the trade off between the manipulability and consensus equilibrium stability;
     \item we also present a rigorous mathematical justification why the large damping yields the feeling of ``sluggish'' in teleoperator systems by using the measure of manipulability;
     \item we illustrate by simulations and intuitive explanations that in a typical industrial/commercial robotic context, the integral action of the low-level PI velocity controller tends to decrease the manipulability of the system.
 \end{enumerate}
  A much general conclusive result obtained is that \emph{networked systems with strong manipulability typically contain a mapping with an infinite gain from the external force to the consensus equilibrium increment, and that the employment of integral action often destroys the strong manipulability of the system (as shown in the two specific cases of the paper)}. This also suggests that achieving consensus of the system does not necessarily implies strong manipulability, and more importantly the formalized concept of manipulability here might act as a general guide for designing controllers for networked systems, especially in the presence of uncertainties. Another contribution of our work is that the proposed scheme achieves the separation of the kinematic and dynamic loops in the case of no task-space velocity measurement and {yields an adaptive kinematic control scheme that can be applied to networked robotic systems with an unmodifiable joint servoing controller yet admitting the design of the joint velocity (or position) command} (e.g., most industrial/commercial robots)}, and this extends the result in \cite{Wang2017_TAC} to address the case of no task-space velocity measurement in the context of multiple robotic systems. A preliminary version of the paper was presented in \cite{Wang2015_CCC}, and the added value of the present paper is the formalization of the concept of manipulability as well as the analysis of the manipulability for networked robotic systems (including the specific teleoperator systems).

\section{Preliminaries}

\subsection{Graph Theory}

Let us give a brief introduction of the graph theory \cite{Godsil2001_Book,Olfati-Saber2004_TAC,Ren2005_TAC,Ren2008_Book} in the scenario that $n$ robotic systems are involved. As is now typically done, we employ a directed graph ${\mathcal G}=({\mathcal V},{\mathcal E})$ to describe the interaction topology among the robotic systems where ${\mathcal V}=\left\{1,2,\dots,n\right\}$ is the vertex set that denotes the collection of the $n$ systems and ${\mathcal E}\subseteq {\mathcal V}\times {\mathcal V}$ is the edge set that denotes the information interaction among the $n$ systems. The set of neighbors of the $i$-th system is denoted by $\mathcal{N}_i=\left\{j|(i,j)\in \mathcal{E}\right\}$. A graph is said to contain a directed spanning tree if there is a vertex $k_0\in \mathcal{V}$ such that any other vertex of the graph has a directed path to $k_0$. The weighted adjacency matrix $\mathcal{W}=\left[w_{ij}\right]$ associated with the graph $\mathcal G$ is defined as $w_{ij}>0$ if $j\in \mathcal{N}_i$, and $w_{ij}=0$ otherwise. In addition, following the standard convention, we make the assumption that $w_{ii}=0$, $\forall i=1,2,\dots,n$. The Laplacian matrix $\mathcal{L}_w=\left[\ell_{w,ij}\right]$ associated with the graph $\mathcal G$ is defined as $\ell_{w,ij}=\Sigma_{k=1}^n w_{ik}$ if $i=j$, and $\ell_{w,ij}=-w_{ij}$ otherwise. Several basic properties associated with the Laplacian matrix $\mathcal{L}_w$ can be described by the following lemma.

\emph{Lemma 1 (\cite{Lin2005_TAC,Ren2005_TAC,Ren2008_Book}):} If $\mathcal{L}_w$ is associated with a directed graph containing a directed spanning tree, then
 \begin{enumerate}
 \item $\mathcal{L}_w$ has a simple zero eigenvalue, and all other eigenvalues of $\mathcal{L}_w$ have positive real parts;
 \item $\mathcal{L}_w$ has a right eigenvector $1_n=\left[1, 1 ,\dots , 1\right]^T$ and a nonnegative left eigenvector $\gamma=\left[\gamma_1, \gamma_2,\dots, \gamma_n\right]^T$ satisfying $\Sigma_{k=1}^n \gamma_k=1$ associated with its zero eigenvalue, i.e., $\mathcal{L}_w 1_n=0$ and $\gamma^T \mathcal{L}_w=0$;
 \item the entry $\gamma_i>0$ if and only if vertex $i$ acts as a root of the graph.
\end{enumerate}

\subsection{Kinematics and Dynamics of Robotic Systems}

Denote by $x_i\in R^m$ the position of the end-effector of the $i$-th robotic system in the task space (e.g., Cartesian space), and its relation with the joint position $q_i\in R^m$ can be written as \cite{Craig2005_Book,Spong2006_Book}
\be
\label{eq1}
x_i=f_i(q_i)
\ee
where $f_i: R^m\to R^m$ denotes a nonlinear mapping.

Differentiating (\ref{eq1}) with respect to time gives the relation between the task-space velocity and joint velocity \cite{Craig2005_Book,Spong2006_Book}
\be
\label{eq2}
\dot x_i=J_i(q_i)\dot q_i
\ee
where $J_i(q_i)\in R^{m\times m}$ is the Jacobian matrix. Since the kinematic parameters are unknown, we cannot obtain the information concerning the task-space position/velocity by the direct kinematics (\ref{eq1}) and (\ref{eq2}). In this paper, we assume that the task-space position $x_i$ is available from the task-space sensors (e.g., a camera) while the task-space velocity $\dot x_i$ is not available. The kinematics (\ref{eq2}) has the linearity-in-parameters property below \cite{Cheah2006_IJRR}.

\emph{Property 1:} The kinematics given by (\ref{eq2}) depends linearly on a constant kinematic parameter vector $\theta_i$, which gives rise to
\be
J_i(q_i)\xi=Z_i(q_i,\xi)\theta_i
\ee
where $\xi\in R^m$ is a vector and $Z_i(q_i,\xi)$ is the kinematic regressor matrix.

The equations of motion of the $i$-th robotic system can be written as \cite{Slotine1991_Book,Spong2006_Book}
\begin{equation}
\label{eq3}
M_i(q_i)\ddot q_i+C_i(q_i,\dot q_i)\dot q_i+g_i(q_i)=\tau_i
\end{equation}
where $M_i \left(q_i\right) \in R^{m\times m}$ is the inertia matrix, $C _i\left( q_i ,\dot
q_i \right) \in R^{m\times m}$ is the Coriolis and centrifugal matrix,
$g_i \left( q_i \right) \in R^m$ is the gravitational torque, and
$\tau_i  \in R^m$ is the joint control torque. Three standard properties associated with the dynamic model (\ref{eq3}) that shall be useful for the controller design and stability analysis are listed as
follows (see, e.g., \cite{Slotine1991_Book,Spong2006_Book}).

\textit{Property 2:} The inertia matrix $M_i (q_i )$ is symmetric and uniformly positive
definite.

\textit{Property 3: }The Coriolis and centrifugal matrix $C_i (q _i,\dot q_i )$ can be
appropriately chosen such that $\dot M_i(q_i) - 2C_i(q_i,\dot q_i)$ is skew-symmetric.

\textit{Property 4: }The dynamics (\ref{eq3}) depends linearly on a
constant dynamic parameter vector $\vartheta_i $, which gives rise to
\begin{equation}
\label{eq4}
M _i\left( q_i  \right)\dot \zeta  + C_i \left( q_i ,\dot q_i \right)\zeta + g _i\left( q_i \right) = Y_i \big( q_i ,\dot q_i
,\zeta ,\dot \zeta \big)\vartheta_i
\end{equation}
where $\zeta \in R^m$ is a differentiable vector,
$\dot {\zeta }$ is the time derivative of $\zeta $, and $Y_i \big( q_i ,\dot q_i
,\zeta ,\dot \zeta \big)$ is the dynamic
regressor matrix.

\section{Adaptive Control and Stabiliy/Manipulability Analysis}

In this section, we investigate the adaptive controller design for the task-space consensus problem of the $n$ robotic systems without involving the task-space velocity measurement and also the manipulability of the closed-loop system. The control objective is to guarantee that the task-space positions of the $n$ robotic systems converge to a common value with their task-space velocities converging to zero, i.e., $x_i-x_j\to 0$ and $\dot x_i\to0$ as $t\to\infty$, $\forall i,j=1,2,\dots,n$.

Let us design a joint reference velocity for the $i$-th system as
\begin{align}
\label{eq6}
\dot q_{r,i}=&\hat J_i^{-1}(q_i)\bigg\{-\Sigma_{j\in\mathcal{N}_i}w_{ij}\left[x_{o,i}-x_{o,j}(t-T_{ij})\right]\nn\\
&-\alpha \int_0^ts_{o,i}^\ast(r)dr\bigg\}
\end{align}
where $\alpha$ is a nonnegative design constant, $T_{ij}$ is the finite constant communication delay from the $j$-th system to the $i$-th system, $\hat J_i(q_i)$ is the estimate of $J_i(q_i)$ (which is obtained by replacing $\theta_i$ in $J_i(q_i)$ with its estimate $\hat\theta_i$), $x_{o,i}$ is the observed quantity of $x_i$ and is updated by the following observer
\begin{align}
\label{eq7}
\dot x_{o,i}=&\hat{J}_i(q_i)\dot q_{r,i}-\beta\left(x_{o,i}-x_i\right)-\lambda\int_0^t[x_{o,i}(r)-x_i(r)]dr\nn\\
=&-\Sigma_{j\in\mathcal{N}_i}w_{ij}\left[x_{o,i}-x_{o,j}(t-T_{ij})\right]\nn\\
&-\alpha \int_0^ts_{o,i}^\ast(r)dr-\beta\left(x_{o,i}-x_i\right)\nn\\
&{ -\lambda\int_0^t[x_{o,i}(r)-x_i(r)]dr}
\end{align}
 where $\beta$ is a positive design constant and $\lambda$ is a nonnegative design constant, and the vector $s_{i}^\ast$ is defined by following \cite{Nuno2011_TAC} as
\be
\label{eq8}
s_{o,i}^\ast=\dot x_{o,i}+\Sigma_{j\in\mathcal{N}_i} w_{ij}\left[x_{o,i}-x_{o,j}(t-T_{ij})\right].
\ee
As is typically done, the signal $x_{o,j}(t-T_{ij})$ in (\ref{eq6}) and (\ref{eq8}) is set as $x_{o,j}(t-T_{ij})\equiv 0$ when $0\le t <T_{ij}$. Note that the proposed observer (\ref{eq7}) is independent of the joint velocity $\dot q_i$, in contrast with the results in \cite{Liu2006_Aut,WangL2014_IJC}, and additionally distributed in the sense that it does not rely on any global information; this distributed observer, also unlike the one in \cite{Mei2012_Aut} that is independent of any physical state information of the system, is coupled to the robotic system by the action $-\beta(x_{o,i}-x_i)-\lambda\int_0^t[x_{o,i}(r)-x_i(r)]dr$. The incorporation of the integral action of $s_{o,i}^\ast$ in (\ref{eq6}) follows the result in \cite{Wang2013_TAC}, and as will be shown later, this integral action allows us to explicitly derive the final consensus value of the systems in the case that $\alpha>0$.

Define a sliding vector
\be
s_i=\dot q_i-\dot q_{r,i}.
\ee
Premultiplying both sides of the above equation by $J_i(q_i)$ and using equation (\ref{eq6}) and Property 1 gives
\begin{align}
\label{eq10}
J_i(q_i)s_i=&\dot x_i+\Sigma_{j\in\mathcal{N}_i}w_{ij}\left[x_{o,i}-x_{o,j}(t-T_{ij})\right]\nn\\
&+\alpha \int_0^ts_{o,i}^\ast(r)dr+Z_i(q_i,\dot q_{r,i})\Delta \theta_i
\end{align}
where $\Delta \theta_i=\hat \theta_i-\theta_i$ is the kinematic parameter estimation error, and in view of (\ref{eq6}), the kinematic regressor matrix $Z_i(q_i,\dot q_{r,i})$ is both adaptive and distributed in that it depends on the estimated kinematic parameter $\hat \theta_i$ updated by the kinematic parameter adaptation law given later and that it does not use any global information of the network.

Subtracting both sides of the kinematics (\ref{eq2}) from those of the observer (\ref{eq7}) and using Property 1, we obtain the closed-loop observer dynamics as
\begin{align}
\label{eq11}
\Delta \dot x_{o,i}=&-\beta \Delta x_{o,i}-\lambda\int_0^t \Delta x_{o,i}(r)dr\nn\\
&+Z_i(q_i,\dot q_{r,i})\Delta \theta_i-J_i(q_i)s_i
\end{align}
where $\Delta x_{o,i}=x_{o,i}-x_i$ denotes the observation error.

Now we propose the control law for the $i$-th system as
\be
\label{eq12}
\tau_i=-K_i s_i+Y_i(q_i,\dot q_i,\dot q_{r,i},\ddot q_{r,i})\hat \vartheta_i
\ee
where $K_i$ is a symmetric positive definite matrix and $\hat \vartheta_i$ is the estimate of $\vartheta_i$. The adaptation laws for updating the estimated parameters $\hat \vartheta_i$  and $\hat \theta_i$ are given as
\begin{align}
\label{eq13}
\dot{\hat \vartheta}_i=&-\Gamma_i Y_i^T (q_i,\dot q_i,\dot q_{r,i},\ddot q_{r,i})s_i\\
\label{eq14}
\dot{\hat \theta}_i=&-\Lambda_i Z_i^T(q_i,\dot q_{r,i})\Delta x_{o,i}
\end{align}
where $\Gamma_i$ and $\Lambda_i$ are both symmetric positive definite matrices. The adaptive control scheme given by (\ref{eq12}), (\ref{eq13}), and (\ref{eq14}) can be considered as an extension of those for a single robotic system in \cite{Wang2017_TAC,Ma1995_ICRA} to the case of multiple robotic systems without involving task-space velocity measurement.  We note that based on (\ref{eq6}), (\ref{eq7}), and (\ref{eq8}), the relation between $\Delta x_{o,i}$ and $s_{o,i}^\ast$ can be expressed as
\be
\label{eq:a2}\beta\Delta x_{o,i}+\lambda\int_0^t\Delta x_{o,i}(r)dr=-\left[s_{o,i}^\ast+\alpha \int_0^t s_{o,i}^\ast(r)dr\right]
\ee
which means that the observation error is also a reflection of the consensus error concerning the observed task-space positions.

\emph{Remark 1:} The use of observed quantities of the task-space positions in the definition of $\dot q_{r,i}$ given by (\ref{eq6}) avoids involving the task-space velocities in the derivative of $\dot q_{r,i}$. This makes the control law (\ref{eq12}) and the dynamic parameter adaptation law (\ref{eq13}) independent of the task-space velocity measurement. Contrary to several observer-based algorithms developed in the context of multiple identical linear systems with the model being exactly known (see, e.g., \cite{Li2010_TCS,Su2014_TIE}), our algorithm considers the more challenging (perhaps more practical) case of nonidentical nonlinear robotic systems with uncertainties. As is well known, the SPR (strictly positive real) condition (or passivity in the case of nonlinear systems) is typically necessary for applying adaptive control, and the robotic system, by reducing its order with sliding vectors, becomes one among such typical nonlinear systems. In addition, the task-space observer (\ref{eq7}) is coupled to the task-space position of the system [by the term $-\beta (x_{o,i}-x_i)-\lambda\int_0^t[x_{o,i}(r)-x_i(r)]dr$ in (\ref{eq7})] and thus in contrast with the one suggested in \cite{Hong2006_Aut,Mei2011_TAC,Wang2014a_WCICA} which is independent of the system's state.

Substituting the control law (\ref{eq12}) into the dynamics (\ref{eq3}) yields
\be
\label{eq15}
M_i(q_i)\dot s_i+C_i(q_i,\dot q_i)s_i=-K_i s_i+Y_i(q_i,\dot q_i,\dot q_{r,i},\ddot q_{r,i})\Delta \vartheta_i
\ee
where $\Delta \vartheta_i=\hat \vartheta_i-\vartheta_i$ is the dynamic parameter estimation error.

The dynamic behavior of the $i$-th robotic system can be described by
\be
\label{eq16}
\begin{cases}
\dot x_{o,i}=-\Sigma_{j\in\mathcal{N}_i}w_{ij}\left[x_{o,i}-x_{o,j}(t-T_{ij})\right]+s_{o,i}^\ast,\\
s_{o,i}^\ast=-\alpha\int_0^t s_{o,i}^\ast(r)dr-\beta \Delta x_{o,i}-\lambda\int_0^t\Delta x_{o,i}(r)dr,\\
\Delta \dot x_{o,i}=-\beta \Delta x_{o,i}-\lambda\int_0^t\Delta x_{o,i}(r)dr+Z_i(q_i,\dot q_{r,i})\Delta \theta_i-J_i(q_i)s_i,\\
\dot{\hat\theta}_i=-\Lambda_i Z_i^T(q_i,\dot q_{r,i})\Delta x_{o,i},\\
M_i(q_i)\dot s_i+C_i(q_i,\dot q_i)s_i=-K_i s_i+Y_i(q_i,\dot q_i,\dot q_{r,i},\ddot q_{r,i})\Delta \vartheta_i,\\
\dot{\hat \vartheta}_i=-\Gamma_i Y_i^T (q_i,\dot q_i,\dot q_{r,i},\ddot q_{r,i})s_i
\end{cases}
\ee
where the upper four equations describe the kinematic loop and the lower two equations the dynamic loop. The interaction between the two loops is reflected in the term $-J_i(q_i)s_i$ in the third equation of (\ref{eq16}).

We are presently ready to formulate the following theorem.

%\emph{Theorem 1:} The control law (\ref{eq12}) and the parameter adaptation laws (\ref{eq13}) and (\ref{eq14}) for the $n$ robotic systems interacting on directed graphs containing a directed spanning tree and subjected to finite constant communication delays ensure 1) the manipulable consensus of the $n$ robotic systems, i.e., $\dot x_i\to 0$ and $x_i-x_j\to 0$ as $t\to\infty$, $\forall i,j=1,2,\dots,n$ in the case that $\alpha=0$, and 2) the scaled weighted average consensus of the $n$ robotic systems, i.e., $x_i\to \left[1/(1+\Sigma_{k=1}^n \Sigma_{l\in{\cal N}_k}\gamma_k w_{kl}T_{kl})\right]\Sigma_{k=1}^n\gamma_k x_{o,k}(0)$ and $\dot x_i\to 0$ as $t\to\infty$, $\forall i=1,2,\dots,n$ in the case that $\alpha>0$.

\emph{Theorem 1:} {If $\lambda>0$, the control law (\ref{eq12}) and the parameter adaptation laws (\ref{eq13}) and (\ref{eq14}) for the $n$ robotic systems interacting on directed graphs containing a directed spanning tree and subjected to finite constant communication delays ensure the manipulable consensus of the $n$ robotic systems, i.e., $\dot x_i\to 0$ and $x_i-x_j\to 0$ as $t\to\infty$, $\forall i,j=1,2,\dots,n$ with $1/\alpha$ acting as the manipulability index, where the manipulable consensus means that the final consensus value can be adjusted by the external stimuli/input exerted at the torque level and the manipulability quantifies the degree of the adjustability of this value. In addition, if $\alpha>0$ and there is no external stimuli/input, the task-space positions of the $n$ robotic systems converge to the scaled weighted average value $\left[1/(1+\Sigma_{k=1}^n \Sigma_{l\in{\cal N}_k}\gamma_k w_{kl}T_{kl})\right]\Sigma_{k=1}^n\gamma_k x_{o,k}(0)$.}

Before presenting the proof of Theorem 1, we first state the following proposition concerning the input-output properties of marginally stable linear systems and it extends the input-output properties (iBIBO stability) stated in \cite{Wang2015_IJC} to the more general case.

{\emph{Proposition 1:} Consider a marginally stable and strictly proper linear time-invariant system $y=G(u)$ with a simple pole at the origin and all other poles in the open left half plane (LHP), where $u\in R^n$ and $y\in R^n$ denote the input and output, respectively. Then
\begin{enumerate}
\item if $\int_0^t u(r)dr\in{\mathcal L}_2$, then $y-\omega(y(0),t)\in {\mathcal L}_2$;
\item if $\int_0^t u(r)dr\in{\mathcal L}_1$, then $y-\omega(y(0),t)\in{\mathcal L}_1$;
\item if $\int_0^t u(r)dr\in{\mathcal L}_\infty$, then $y\in{\mathcal L}_\infty$
\end{enumerate}
where $\omega(y(0),t)$ is the zero-input response of the system and is a bounded function in terms of time and the initial value $y(0)$.}

\emph{Proof:} The proof follows similar procedures as in \cite{Wang2015_IJC}. The representation of the system in frequency domain can be written as $Y(p)=G(p)[U(p)+F(y(0))]$ with $p$ denoting the Laplace variable and $Y(p)$ and $U(p)$ the Laplace transforms of $y$ and $u$, respectively. We now rewrite the system as $Y(p)=pG(p)[U(p)/p]+G(p)F(y(0))$ where $pG(p)$ is a biproper function \cite{Ioannou1996_Book} and $U(p)/p$, as is well known, is the Laplace transform of $\int_0^t u(r)dr$. The second part of $Y(p)$ denoted by $\omega(y(0),t)=G(p)F(y(0))$ is obviously bounded and additionally converges to some constant vector, according to the standard linear system theory. Furthermore, we have that the transfer function $pG(p)$ is exponentially stable since the simple zero pole of $G(p)$ is cancelled by the factor $p$ \cite{Desoer1975_Book}. Then following similar analysis as in the proof of Corollary 3.3.2 in \cite{Ioannou1996_Book}, we obtain that the time-domain counterpart of $pG(p)[U(p)/p]$ (i.e., $y(t)-\omega(y(0),t)$) is square-integrable if $\int_0^t u(r)dr\in{\mathcal L}_2$. The conclusions 2) and 3) can be similarly derived.  \hfill {\small $\blacksquare$}

\emph{Proof of Theorem 1:} Following the standard practice (see, e.g., \cite{Slotine1987_IJRR,Ortega1989_Aut}), we consider the Lyapunov-like function candidate for the fifth and sixth subsystems in (\ref{eq16}) $
V_i=(1/2) s_i^T M_i(q_i)s_i+(1/2) \Delta \vartheta_i^T \Gamma_i^{-1}\Delta \vartheta_i
$,
and by exploiting {Property 3}, we obtain the time derivative of $V_i$ as
$
\dot V_i=-s_i^T K_i s_i\le 0
$, which yields the result that $s_i\in {\cal L}_2\cap {\cal L}_\infty$ and $\hat \vartheta_i\in {\cal L}_\infty$, $\forall i$.

By using the well-known fact that $J_i(q_i)$ is bounded, we obtain the result that $J_i(q_i)s_i\in {\cal L}_2$, and consequently, there exists a positive constant $l_{M,i}$ such that $\int_0^t s_i^T(r) J_i^T(q_i(r))J_i(q_i(r))s_i(r) dr\le l_{M,i}$ for all $t\ge 0$, $\forall i$. Then, we consider the following quasi-Lyapunov function candidate
\begin{align}
\label{eq19}
V_i^\ast=&\frac{1}{2}\Delta x_{o,i}^T\Delta x_{o,i}+\frac{\lambda}{2}\left[\int_0^t \Delta x_{o,i}(r)dr\right]^T\left[\int_0^t \Delta x_{o,i}(r)dr\right]
\nn\\&+\frac{1}{2\beta}
\bigg[l_{M,i}-\int_0^t s_i^T(r) J_i^T(q_i(r))J_i(q_i(r))s_i(r) dr\bigg]\nn\\
&+\frac{1}{2}\Delta \theta_i^T \Lambda_i^{-1}\Delta \theta_i
\end{align}
where the adoption of the third term in $V_i^\ast$ follows the typical practice (see, e.g., \cite[p.~118]{Lozano2000_Book}) and is for taking into account the interaction between the kinematic and dynamic loops, $\forall i$. The time derivative of $V_i^\ast$ along the third and fourth subsystems in (\ref{eq16}) can be shown to satisfy
\begin{align}
\label{eq21}
\dot V_i^\ast\le&-\frac{\beta}{2}\Delta x_{o,i}^T\Delta x_{o,i}\le 0, \forall i
\end{align}
where we have used the following result that is derived from the standard basic inequalities
\begin{align*}
-\Delta x_{o,i}^T J_i(q_i)s_i\le& \frac{\beta}{2} \Delta x_{o,i}^T \Delta x_{o,i}+\frac{1}{2\beta}s_i^T J_i^T(q_i)J_i(q_i)s_i.
\end{align*}
 The inequality (\ref{eq21}) as well as the definition of $V_i^\ast$ given by (\ref{eq19}) immediately leads us to obtain that $\Delta x_{o,i}\in {\mathcal L}_2\cap{\mathcal L}_\infty$ and $\hat{\theta}_i\in {\mathcal L}_\infty$, and that $\int_0^t \Delta x_{o,i}(r)dr\in {\mathcal L}_\infty$ if $\lambda>0$, $\forall i$.

From the second subsystem in (\ref{eq16}), we obtain by the Laplace transformation that
$$S_{o,i}^\ast(p)=-(\beta p+\lambda)/(p+\alpha)\Delta X_{o,i}(p),$$
and this leads us to obtain that $s_{o,i}^\ast\in {\mathcal L}_2\cap{\mathcal L}_\infty$ according to the input-output properties of biproper linear systems \cite[p.~82]{Ioannou1996_Book}, $\forall i$. Applying Laplace transformation to the first subsystem in (\ref{eq16}) yields
 \be
 pX_{o,i}(p)-x_{o,i}(0)=-\Sigma_{j\in{\mathcal N}_i}w_{ij}[X_{o,i}(p)-e^{-T_{ij}p}X_{o,j}(p)]+S_{o,i}^\ast(p).
 \ee
By letting $\Phi_i(p)=pX_{o,i}(p)-x_{o,i}(0)$ denote the Laplace transform of $\dot x_{o,i}$, $\forall i$, we have that
\begin{align}
\Phi_i(p)=&-\Sigma_{j\in{\mathcal N}_i}w_{ij}\frac{\Phi_i(p)-e^{-T_{ij}p}\Phi_j(p)}{p}\nn\\
&-\Sigma_{j\in{\mathcal N}_i}w_{ij}\frac{x_{o,i}(0)- e^{-T_{ij}p}x_{o,j}(0)}{p}+S_{o,i}^{\ast}(p).
\end{align}
Stacking up all the equations like above with further manipulations gives
\begin{align}
\label{eq:a1}
\Phi(p)=[G(p)\otimes I_m]\{-[({\mathcal D}_w-{\mathcal W}_T(p))\otimes I_m]x_o(0)+pS_o^\ast(p)]
\end{align}
where ${\mathcal D}_w={\rm diag}[\Sigma_{j\in{\mathcal N}_i}w_{ij},i=1,\dots,n]$, ${\mathcal W}_T(p)=[w_{ij}e^{-T_{ij}p}]$, $G(p)=(pI_n+{\mathcal D}_w-{\mathcal W}_T(p))^{-1}$, $\otimes$ denotes the Kronecker product \cite{Brewer1978_TCS}, $x_o=[x_{o,1}^T,\dots,x_{o,n}^T]^T$, $S_o^\ast(p)=[S_{o,1}^{\ast T}(p),\dots,S_{o,n}^{\ast T}(p)]^T$, and $\Phi(p)=[\Phi_1^T(p),\dots,\Phi_n^T(p)]^T$. The time-domain counterpart of $pS_o^\ast (p)$ can obviously be written as $\dot s_o^\ast(t)+\delta(t)s_o^\ast(0)$ where $s_o^\ast(t)=\left[s_{o,1}^{\ast T}(t),\dots,s_{o,n}^{\ast T}(t)\right]^T$$\delta(t)$ denotes the standard Dirac delta function, and the integral of this function is $\int_0^t [\dot s_o^\ast(r)+\delta(r)s_o^\ast(0)]dr=s_o^\ast(t)\in{\mathcal L}_2\cap{\mathcal L}_\infty$.
 From \cite{Nuno2011_TAC}, we know that $G(p)$ has a simple pole at the origin and its rest poles are in the open LHP. Therefore, from Proposition 1 and (\ref{eq:a1}), we immediately obtain that $\dot x_o\in {\mathcal L}_\infty$ and $\dot x_o-\omega_1(x_o(0),t)\in{\mathcal L}_2$ with $\omega_1(x_o(0),t)$ being the time-domain counterpart of $-[G(p)\otimes I_m][({\mathcal D}_w-{\mathcal W}_T(p))\otimes I_m]x_o(0)$. It can be shown that $\omega_1(x_o(0),t)\to 0$ as $t\to\infty$ \cite{Wang2014_TAC} (this convergence is exponential according to the standard linear system theory) and obviously $\omega_1(x_o(0),t)\in{\mathcal L}_2$, which implies that $\dot x_o\in{\mathcal L}_2$.
 This directly gives the result that $\Sigma_{j\in{\mathcal N}_i}w_{ij}[x_{o,i}-x_{o,j}(t-T_{ij})]\in{\mathcal L}_2\cap{\mathcal L}_{\infty}$, $\forall i$. From (\ref{eq6}), we obtain that $\dot q_{r,i}\in{\mathcal L}_\infty$ if $\hat J_i(q_i)$ is nonsingular, and thus $\dot q_i\in{\mathcal L}_\infty$, $\forall i$. From (\ref{eq2}), we obtain that $\dot x_i\in{\mathcal L}_\infty$, $\forall i$. By the differentiation of (\ref{eq7}), we obtain that $\ddot x_{o,i}\in{\mathcal L}_\infty$ and therefore  $\dot s_{o,i}^\ast\in{\mathcal L}_\infty$, which implies that $\dot x_{o,i}$ and $s_{o,i}^\ast$ are both uniformly continuous, $\forall i$. According to the properties of square-integrable and uniformly continuous functions \cite[p.~232]{Desoer1975_Book}, we obtain that $\dot x_{o,i}\to 0$ and $s_{o,i}^\ast\to 0$ as $t\to\infty$, $\forall i$. From the first subsystem in (\ref{eq16}), we immediately obtain that $\Sigma_{j\in{\mathcal N}_i}w_{ij}[x_{o,i}-x_{o,j}(t-T_{ij})]\to 0$ as $t\to\infty$, $\forall i$. Considering the standard fact that $x_{o,j}(t)-x_{o,j}(t-T_{ij})=\int_0^{T_{ij}}\dot x_{o,j}(t-r)dr \to 0$ as $t\to\infty$, $\forall j\in{\mathcal N}_i$, $\forall i$, we then obtain that $\Sigma_{j\in{\mathcal N}_i}w_{ij}[x_{o,i}-x_{o,j}]\to 0$ as $t\to\infty$, $\forall i$, which directly yields the result that $({\mathcal L}_w\otimes I_m)x_o\to 0$ as $t\to\infty$. From Lemma 1, we obtain that $x_{o,i}-x_{o,j}\to 0$ as $t\to\infty$, $\forall i,j$. The result that $\dot x_{o,i}\in{\mathcal L}_\infty$ and $\dot x_i\in{\mathcal L}_\infty$ implies that $\Delta x_{o,i}\in{\mathcal L}_\infty$, which means that $\Delta x_{o,i}$ is uniformly continuous, $\forall i$. We then obtain that $\Delta x_{o,i}\to 0$ as $t\to\infty$ from the properties of square-integrable and uniformly continuous functions \cite[p.~232]{Desoer1975_Book}, $\forall i$. Hence, we have that $x_i-x_j\to 0$ as $t\to\infty$, $\forall i,j$.

  From (\ref{eq14}), we obtain that $\dot{\hat \theta}_i\in{\mathcal L}_\infty$, which implies that $\dot{\hat J}_i(q_i)$ is bounded, $\forall i$. This leads us to obtain that $\ddot q_{r,i}\in{\mathcal L}_\infty$, $\forall i$. From (\ref{eq15}) and exploiting Property 2, we obtain that $\dot s_i\in {\cal L}_\infty$ and further $\ddot q_i\in{\cal L}_\infty$, $\forall i$. Based on the differentiation of (\ref{eq2}), i.e., $\ddot x_i=J_i(q_i)\ddot q_i+\dot J_i(q_i)\dot q_i$, we have that $\ddot x_i\in {\cal L}_\infty$ and thus $\Delta \ddot x_{o,i}\in{\mathcal L}_\infty$, which implies that $\Delta \dot x_{o,i}$ is uniformly continuous, $\forall i$. From Barbalat's Lemma \cite{Slotine1991_Book}, we have that $\Delta \dot x_{o,i}\to 0$ as $t\to\infty$ and thus $\dot x_i\to 0$ as $t\to\infty$, $\forall i$.

In the case that $\alpha>0$, consider the following system
\begin{align}
\label{eq22}
\dot x_{o,i}=&-\Sigma_{j\in{\mathcal N}_i} w_{ij}\left[x_{o,i}-x_{o,j}(t-T_{ij})\right]\nn\\
&+s_{o,i}^\ast, \text{ } i=1,2,\dots,n.
\end{align}
First, we have from \cite{Wang2014_TAC} that the system (\ref{eq22}) with $s_{o,i}^\ast$, $i=1,2,\dots,n$ as the input and $x_{o,i}$, $i=1,2,\dots,n$ as the output is integral-bounded-input bounded-output (iBIBO) stable in the sense of \cite{Wang2015_IJC} and thus we obtain that $x_{o,i}\in {\cal L}_\infty$ since $\int_0^t s_{o,i}^\ast(r)dr\in {\cal L}_\infty$ from the second subsystem of (\ref{eq16}), $\forall i$. This gives rise to the consequence that $x_i\in{\mathcal L}_\infty$ since $\Delta x_{o,i}\in{\mathcal L}_\infty$, $\forall i$.

{We next rigorously show that $1/\alpha$ measures the manipulability of the system responding to an external input or stimuli exerted at the torque level, mainly by resorting to the standard concepts and analysis approaches concerning input-output stability (see, e.g., \cite{Desoer1975_Book,Schaft2000_Book,Vidyasagar1993_Book,Khalil2002_Book,Ioannou1996_Book,Rotea1993_AUT,Sontag1998_SCL,Angeli2000_TAC}). As is known (see, e.g., \cite{Nuno2011_TAC,Wang2014_TAC}), the final consensus value depends linearly on the integral of $s_{o,i}^\ast$, $i=1,\dots,n$. Suppose that an external input $\tau_{h,i}$ is exerted on the $i$-th system and that the $i$-th system acts as the root of the graph, which gives
\be
\label{eq23}
\begin{cases}
\dot x_{o,i}=-\Sigma_{j\in\mathcal{N}_i}w_{ij}\left[x_{o,i}-x_{o,j}(t-T_{ij})\right]+s_{o,i}^\ast,\\
s_{o,i}^\ast=-\alpha\int_0^t s_{o,i}^\ast(r)dr-\beta \Delta x_{o,i}-\lambda\int_0^t\Delta x_{o,i}(r)dr,\\
\Delta \dot x_{o,i}=-\beta \Delta x_{o,i}-\lambda\int_0^t\Delta x_{o,i}(r)dr+Z_i(q_i,\dot q_{r,i})\Delta \theta_i-J_i(q_i)s_i,\\
\dot{\hat\theta}_i=-\Lambda_i Z_i^T(q_i,\dot q_{r,i})\Delta x_{o,i},\\
M_i(q_i)\dot s_i+C_i(q_i,\dot q_i)s_i=-K_i s_i+Y_i(q_i,\dot q_i,\dot q_{r,i},\ddot q_{r,i})\Delta \vartheta_i+\tau_{h,i},\\
\dot{\hat \vartheta}_i=-\Gamma_i Y_i^T (q_i,\dot q_i,\dot q_{r,i},\ddot q_{r,i})s_i.
\end{cases}
\ee
Considering the same Lyapunov-like function candidate as before for the lower two subsystems yields $\dot V_i\le-\frac{1}{2}s_i^T K_i s_i+\frac{1}{2}\tau_{h,i}^T K_i^{-1}\tau_{h,i}$ (by resorting to the standard basic inequalities), and this leads us to obtain that the $\mathcal L_2$-gain from $\tau_{h,i}$ to $s_i$ is not greater than $1/\lambda_{\min}\{K_i\}$ where $\lambda_{\min}\{\cdot\}$ denotes the minimum eigenvalue of a matrix.  Consider the third and fourth subsystems of (\ref{eq23}) and using the same quasi-Lyapunov function candidate as before yields $\dot V_i^\ast\le -\frac{\beta}{2}\Delta x_{o,i}^T\Delta x_{o,i}\le 0$. Then we obtain that the $\mathcal L_2$-gain from $J_i(q_i)s_i$ to $\Delta x_{o,i}$ is not greater than $1/\beta$, and in addition the square-integrability of $J_i(q_i)s_i$ yields the boundedness of $\int_0^t \Delta x_{o,i}(r)dr$ with the $\mathcal L_{2}$$\mapsto$${\mathcal L_\infty}$-gain (see, e.g., \cite{Rotea1993_AUT,Sontag1998_SCL,Angeli2000_TAC}) from $J_i(q_i)s_i$ to $\int_0^t \Delta x_{o,i}(r)dr$ being not greater than $1/\sqrt{\beta\lambda}$. Consider the second subsystem in (\ref{eq23}) with $\int_0^t s_{o,i}^\ast(r)dr$ as the output, which can be considered as the composite of that generated by the input $\Delta x_{o,i}$ and that generated by the input $\int_0^t \Delta x_{o,i}(r)dr$. For this subsystem, we obtain that the $\mathcal L_2$-gain from $\Delta x_{o,i}$ to $\int_0^t s_{o,i}^\ast(r)dr$ is $\beta/\alpha$ and the $\mathcal L_\infty$-gain from $\int_0^t \Delta x_{o,i}(r)dr$ to $\int_0^t s_{o,i}^\ast(r)dr$ is $\lambda/\alpha$, by following the typical practice (see, e.g., \cite{Vidyasagar1993_Book,Desoer1975_Book}). Hence $\int_0^t s_{o,i}^\ast(r)dr$ can be written as two parts, namely, $\int_0^t s_{o,i}^\ast(r)dr=\Pi_1+\Pi_2$ with $\Pi_1$ due to the input $\Delta x_{o,i}$ and $\Pi_2$ due to the input $\int_0^t \Delta x_{o,i}(r)dr$. The $\mathcal L_2$-gain from $\tau_{h,i}$ to $\Pi_1$ is not greater than $$\frac{\sqrt{\sup_t\lambda_{\max}\{J_i^T(q_i) J_i(q_i)\}}}{\alpha\lambda_{\min}\{K_i\}}$$
with $\lambda_{\max}\{\cdot\}$ denoting the maximum eigenvalue of a matrix, and the $\mathcal L_2$$\mapsto$$\mathcal L_\infty$-gain from $\tau_{h,i}$ to $\Pi_2$ is not greater than
$$\frac{\sqrt{\sup_t\lambda_{\max}\{J_i^T(q_i) J_i(q_i)\}\lambda}}{\alpha\lambda_{\min}\{K_i\}\sqrt{\beta}}.$$
The contribution of $\int_0^t s_{o,i}^\ast(r)dr$ to the observed consensus equilibrium increment $\Sigma_{k=1}^n \gamma_k [x_{o,k}-x_{o,k}(0)]$ can be described by the gain $c^\ast\gamma_i$ with $c^\ast$ being a positive constant (this can be derived by following the standard practice based on the result in \cite{Wang2014_TAC}). On the other hand, the consensus equilibrium increment can be written as
\begin{align}
\Sigma_{k=1}^n &\gamma_k [x_{k}-x_{k}(0)]\nn\\
=&\underbrace{\Sigma_{k=1}^n \gamma_k [x_{o,k}-x_{o,k}(0)]}_\text{observed equilibrium increment}\nn\\
&+\Sigma_{k=1}^n\gamma_k [-\Delta x_{o,k}+\Delta x_{o,k}(0)].
\end{align}
Hence, the $\mathcal L_2$-gain from $\tau_{h,i}$ to the 1st portion of the consensus equilibrium increment can be shown to satisfy
\begin{align}
&{\mu}_{\mathcal L_2\mapsto \mathcal L_2}\nn\\
&\le\gamma_i\left[\frac{c^\ast\sqrt{\sup_t\lambda_{\max}\{J_i^T(q_i) J_i(q_i)\}}}{\alpha\lambda_{\min}\{K_i\}}\right.\nn\\
&\left.\qquad+\frac{\sqrt{\sup_t\lambda_{\max}\{J_i^T (q_i)J_i(q_i)\}}}{\beta\lambda_{\min}\{K_i\}}\right],
\end{align}
and the $\mathcal L_2$$\mapsto$$\mathcal L_\infty$-gain from $\tau_{h,i}$ to the 2nd portion of the consensus equilibrium increment can be shown to satisfy
\begin{align}&\mu_{\mathcal L_2\mapsto \mathcal L_\infty}\nn\\
&\le\gamma_i\left[\frac{c^\ast\sqrt{\sup_t\lambda_{\max}\{J_i^T(q_i) J_i(q_i)\}\lambda}}{\alpha\lambda_{\min}\{K_i\}\sqrt{\beta}}\right.\nn\\
&\left.\qquad+\frac{\sqrt{\sup_t\lambda_{\max}\{J_i^T(q_i) J_i(q_i)\}}}{\sqrt{\beta}\lambda_{\min}\{K_i\}}\right].
\end{align}
 In particular, these two gains associated with the mapping (one is concerning $\mathcal L_2$ to $\mathcal L_2$ and the other $\mathcal L_2$ to $\mathcal L_\infty$) become infinite in the case that $\alpha=0$ since a pure integral operation is involved in the mapping whose $\mathcal L_2$-gain (or $\mathcal H_\infty$ norm of the corresponding transfer function) and $\mathcal L_\infty$-gain, as is well known, are infinite. Therefore, the value of $1/\alpha$ is indeed a qualified measure of the manipulability of the system, and an immediate conclusion is that setting $\alpha=0$ implies the infinite manipulability.}

In the case that $\alpha>0$ and there is no external stimuli/input, we obtain from the third subsystem of (\ref{eq16}) that $-\lambda\int_0^t \Delta x_{o,i}(r)dr+Z_i(q_i,\dot q_{r,i})\Delta\theta_i\to 0$ as $t\to\infty$, $\forall i$. In accordance with the properties of square-integrable and uniformly continuous functions \cite[p.~232]{Desoer1975_Book}, we obtain that $s_i\to 0$ and thus $\dot q_{r,i}\to \dot q_i$ as $t\to\infty$, $\forall i$. Considering the fact that $\dot q_{r,i}$ given by (\ref{eq6}) can be rewritten as [using (\ref{eq7})]
\be
\dot q_{r,i}=\hat J_i^{-1}(q_i)\left[\dot x_{o,i}+\beta\Delta x_{o,i}+\lambda\int_0^t\Delta x_{o,i}(r)dr\right]
\ee
and that $\dot x_{o,i}\to 0$ and $\Delta x_{o,i}\to 0$ as $t\to\infty$, we obtain that
$\lambda J_i(q_i)\hat J_i^{-1}(q_i)\int_0^t\Delta x_{o,i}(r)dr\to 0$ as $t\to\infty$ since $\dot x_i\to 0$ as $t\to\infty$, $\forall i$. If $\hat J_i(q_i)$ and $J_i(q_i)$ are nonsingular, we obtain that $\int_0^t\Delta x_{o,i}(r)dr\to 0$, $\forall i$. From the second subsystem of (\ref{eq16}), we can directly obtain by the standard final value theorem that $\int_0^ts_{o,i}^\ast(r)dr\to 0$ as $t\to\infty$, $\forall i$. Then using the result in \cite{Wang2014_TAC}, we obtain from (\ref{eq22}) that $x_{o,i}\to \left[1/(1+\Sigma_{k=1}^n \Sigma_{l\in{\cal N}_k}\gamma_k w_{kl}T_{kl})\right]\Sigma_{k=1}^n\gamma_k x_{o,k}(0)$ and further that $x_{i}\to \left[1/(1+\Sigma_{k=1}^n \Sigma_{l\in{\cal N}_k}\gamma_k w_{kl}T_{kl})\right]\Sigma_{k=1}^n\gamma_k x_{o,k}(0)$ as $t\to\infty$, $\forall i$.
\hfill {\small $\blacksquare$}

\emph{Remark 2:} If we set $\lambda=0$ [i.e., removing the integral action of the observation error in (\ref{eq7})] and $\alpha>0$, it can also be shown that the task-space positions of the robotic systems converge to the scaled weighted average value $\left[1/(1+\Sigma_{k=1}^n \Sigma_{l\in{\cal N}_k}\gamma_k w_{kl}T_{kl})\right]\Sigma_{k=1}^n\gamma_k x_{o,k}(0)$, by following the proof of Theorem 1 in \cite{Wang2014_TAC}. Nevertheless, the asymptotic consensus of the systems can possibly no longer be maintained under an external stimuli [e.g., a task-space PD(proportional-derivative)-like input] due to the absence of the integral action of the observation error, and in fact there would generally exist a steady-state observation error. On the other hand, if we set $\lambda=\alpha=0$, the asymptotic consensus of the systems under an external task-space PD-like stimuli can be recovered, but we can no longer ensure that the task-space positions of the robotic systems converge to a constant (or bounded) value. Furthermore, {our result relies on the condition that the estimated Jacobian matrix $\hat J_i(q_i)$ is nonsingular in the kinematic parameter adaptation process, $\forall i$. This can be ensured by the assumption that the manipulator is away from the singular configuration and by the use of the parameter projection algorithms (see, e.g., \cite{Cheah2006_IJRR,Cheah2006_TAC,Dixon2007_TAC})}.

\emph{Remark 3:} The proposed adaptive control scheme requires the communication between the robotic systems. In this context, one might attempt to employ the existing distributed-observer-based schemes (e.g., \cite{Hong2006_Aut,Mei2012_Aut,Wang2014a_WCICA}) where the consensus of the communicated quantities is completely separated from the dynamics of the robotic systems and the objective of each robotic system is to unidirectionally track the corresponding communicated quantity. The main limitations of this strategy may lie in the fact that each robotic system is unidirectionally coupled to the virtual consensus system, giving rise to the consequence that the virtual consensus system is independent of the robotic systems and that the robotic systems are actually not coupled with each other. This renders the consensus behavior unresponsive to external physical commands (e.g., in the scenario that one robotic system is tuned to a static position by a human operator, the other systems, however, will not yield any tendency of trying to achieve consensus with this system since each of them is actually only coupled to the artificial communicated quantity). %Furthermore, this strategy cannot naturally give rise to the separation of the kinematic and dynamic loops of the robotic systems and the avoidance of the task-space velocity measurement.
The proposed distributed-observer-based adaptive control, by introducing a feedback coupling action $-\beta(x_{o,i}-x_i)-\lambda\int_0^t [x_{o,i}(r)-x_i(r)]dr$, results in the manipulable consensus of the robotic systems (i.e., the consensus behavior is responsive to external physical manipulation), in contrast with the unresponsive behavior of the existing distributed-observer-based algorithms mentioned above. %, and simultaneously avoids the task-space velocity measurement.
An important application of the manipulable consensus is bilateral teleoperation involving two manipulators, i.e., the master and slave where the master is typically manipulated by a human operator and the slave tries to maintain consensus with the master (see, e.g., \cite{Hokayem2006_AUT}).

\emph{Remark 4:} {Most leader-based schemes and general consensus schemes without using the distributed observers as in \cite{Hong2006_Aut,Mei2012_Aut,Wang2014a_WCICA} in the literature can also achieve the manipulable consensus introduced here, e.g., the consensus schemes for identical single-integrator systems in \cite{Olfati-Saber2004_TAC} and the consensus schemes without or with the integral action of the sliding vector (defined as the weighted sum of the velocity and neighbor-to-neighbor position consensus errors) for uncertain nonidentical Euler-Lagrange systems or robotic systems (e.g., \cite{Nuno2011_TAC,Wang2014_TAC,Wang2013_TAC}). In our opinion, the explicit-physical-leader-based schemes can at best be unilaterally manipulable (i.e., the maneuvering direction is unilaterally from the physical leader to the followers and no reflection force feedback to the human operator) while it seems hard to justify the discussion of manipulability of virtual-leader-based schemes. The main contribution of the study here is to formalize the concept of manipulability and show that the gain of the integral action concerning the sliding vector acts as a qualified measure of manipulability of the closed-loop system, and to address the case of no task-space velocity measurement by developing a new task-space observer which is more robust than the existing ones.} More generally, our result presents the rigorous mathematical explanation behind the operation of many networked systems (including the teleoperator systems), and in fact the existence of a mapping with an infinite gain from the external force/torque to the consensus equilibrium increment is actually the fundamental condition that allows the human operator to easily maneuver the system (without too much fatigue).

\emph{Remark 5:} We may revisit the teleoperator system involving two robotic systems under a PD controller with gravitation compensation (see, e.g., \cite{Lee2006_TRO,Nuno2009_IJRR})
\be
\begin{cases}
M_1(q_1)\ddot q_1+C_1(q_1,\dot q_1)\dot q_1=-K_D\dot q_1-K_P(q_1-q_2)+\tau_h\\
M_2(q_2)\ddot q_2+C_2(q_2,\dot q_2)\dot q_2=-K_D\dot q_2-K_P(q_2-q_1)
\end{cases}
\ee
where $K_D$ and $K_P$ are diagonal positive definite matrices and $\tau_h$ is the torque exerted by the human operator. The fundamental issue here is why the human operator can adjust the consensus equilibrium without using so much efforts. To this end, similarly to the previous discussions and analysis in the proof of Theorem 1, we consider the mapping from $\tau_h$ to the consensus equilibrium increment and analyze the $\mathcal L_2$-gain of this mapping. Consider the Lyapunov function candidate (see, e.g., \cite{Nuno2009_IJRR,Lee2006_TRO})
\begin{align}
V=&(1/2)\dot q_1^T M_1(q_1)\dot q_1+(1/2)\dot q_2^T M_2(q_2)\dot q_2\nn\\
&+(1/2)(q_1-q_2)^T K_P (q_1-q_2),
\end{align}
and this yields (using Property 3)
\begin{align}
\dot V=&-\dot q_1^T K_D \dot q_1-\dot q_2^T K_D \dot q_2+\dot q_1^T \tau_h\nn\\
\le &-\frac{1}{2}\dot q_1^T K_D \dot q_1-\dot q_2^T K_D \dot q_2+\frac{1}{2}\tau_h^T K_D^{-1}\tau_h
\end{align}
where we have exploited the following result obtained by the standard basic inequalities
\be
\dot q_1^T \tau_h\le \frac{1}{2}\dot q_1^T K_D \dot q_1+\frac{1}{2}\tau_h^T K_D^{-1}\tau_h.
\ee
Following similar arguments as in the proof of Theorem 1, we obtain that the $\mathcal L_2$-gain from $\tau_h$ to $\dot q_1$ is not greater than $1/\lambda_{\min}\{K_D\}$ and that from $\tau_h$ to $\dot q_2$ is not greater than $1/(\sqrt{2}\lambda_{\min}\{K_D\})$.
%
%In the case that $q_1=q_2=q^\ast$ (consensus is reached), we obtain that
%\begin{align}
%&[M_1(q^\ast)+M_2(q^\ast)]\ddot q^\ast+[C_1(q^\ast,\dot q^\ast)+C_2(q^\ast,\dot q^\ast)]\dot q^\ast\nn\\
%&=-2K_D \dot q^\ast+\tau_h.
%\end{align}
%
%which defines a bounded mapping from $\tau_h$ to $\dot q^\ast$ with the gain relation of the two quantities being inversely proportional to $K_D$ (this can be shown by the standard practice---see, e.g., \cite{Khalil2002_Book}).
Let the consensus equilibrium be represented by $q_c=\frac{q_1+q_2}{2}$. Then the $\mathcal L_2$-gain from $\tau_h$ to $\dot q_c$ is not greater than $$\left(\frac{1+\sqrt{2}}{2\sqrt{2}}\right)\frac{1}{\lambda_{\min}\{K_D\}}.$$
On the other hand, the mapping from $\dot q_c$ to $q_c-q_c(0)$ can be represented in frequency domain by $1/p$ whose $\mathcal H_\infty$ norm (i.e., the $\mathcal L_2$-gain of the mapping), as is well known, is $\sup_{\omega^\ast}\frac{1}{|\sqrt{-1}\omega^\ast|}=\infty$. Overall, the gain of the mapping from $\tau_h$ to the consensus equilibrium increment $q_c-q_c(0)$ is infinite and can be specifically considered as the composite of a pure integral operation and a mapping with the upper bound of its finite $\mathcal L_2$-gain being inversely proportional to the damping. In the literature on teleoperation, it is well recognized that physically and intuitively, large damping would give rise to the feeling of ``sluggish''; we here present a rigorous mathematical justification of this ``sluggish'' feeling, i.e., the measure of manipulability excluding the pure integral component (whose upper bound would be inversely proportional to the damping) becomes decreased with a large damping.

\section{Adaptive Kinematic Control}

In most industrial/commercial robots, the available design input is the joint velocity (or position) rather than the joint torque and in fact the torque control is typically hidden from the user (see, e.g., \cite{Grotjahn2002_IJRR}). The existing task-space consensus controllers (e.g., \cite{WangL2014_IJC,Liu2015_TMECH}) cannot be applied to this category of robots due to the dependence on the modification of the low-level feedback controller architecture. The separation property of the proposed dynamic controller in Sec. III allows us to conveniently obtain an adaptive kinematic controller that is applicable to robots having an unmodifiable joint servoing module yet admitting the design of the joint velocity (or position) command. Specifically, we have the following theorem.

\emph{Theorem 2:} Suppose that the low-level joint controllers for the $n$ robotic systems can ensure that the joint velocity tracking error $s_i=\dot q_i-\dot q_{r,i}\in {\mathcal L}_2\cap{\mathcal L}_\infty$, $\forall i=1,2,\dots,n$ and $\lambda>0$, and that there is no task-space and joint-space velocity measurement. Then the adaptive kinematic controller given by (\ref{eq7}), (\ref{eq6}), and (\ref{eq14}) for the $n$ robotic systems on directed graphs containing a directed spanning tree ensures the manipulable consensus of the $n$ robotic systems, i.e., $\dot x_i\to 0$ and $x_i-x_j\to 0$ as $t\to\infty$, $\forall i,j=1,2,\dots,n$ with $1/\alpha$ acting as the manipulability index. In addition, if $\alpha>0$ and there is no external stimuli/input, the task-space positions of the $n$ robotic systems converge to the scaled weighted average value $\left[1/(1+\Sigma_{k=1}^n \Sigma_{l\in{\cal N}_k}\gamma_k w_{kl}T_{kl})\right]\Sigma_{k=1}^n\gamma_k x_{o,k}(0)$.

\emph{Proof:}  The proof can be completed by following similar steps as in the proof of Theorem 1. In fact, choose $V_i^\ast$ in (\ref{eq19}) as the quasi-Lyapunov function candidate and it can be directly shown that the derivative of $V_i^\ast$ along the trajectories of the third and fourth subsystems in (\ref{eq16}) satisfies the inequality
$
\dot V_i^\ast\le -(\beta/2)\Delta x_{o,i}^T \Delta x_{o,i}\le 0.
$
Then, it can be shown that the manipulable consensus of the $n$ robotic systems is indeed realized, using a procedure similar to that in the proof of Theorem 1. \hfill {\small $\blacksquare$}

\emph{Remark 6:} The result given in Theorem 2 demonstrates that for robotic systems having an unmodifiable joint servoing controller yet allowing the design of the joint velocity (or position) command (e.g., most industrial/commercial robots), it is also possible to achieve consensus in the presence of kinematic uncertainties and absence of {task-space and joint-space velocity measurement} and {without directly involving task-space position measurement (implying enhanced robustness since $x_{o,i}$ can be considered as a filtered quantity of $x_i$ and $\hat \theta_i$ as a quantity yielded by an integration concerning $x_i$)}. The adaptive kinematic controller here, in contrast with \cite{Wang2017_TAC}, does not directly involve the task-space position due to the introduced new task-space observer. In addition, similar to \cite{Wang2017_TAC}, certain module-like properties of the proposed adaptive kinematic controller are guaranteed in the sense that the joint servoing controller is merely demanded to guarantee the square-integrability and boundedness of the joint velocity tracking error (i.e., $s_i\in{\mathcal L}_2\cap{\mathcal L}_\infty$, $\forall i$). Obviously, the adaptive joint servoing controller given by (\ref{eq12}) and (\ref{eq13}) is only a special case. The requirement that the joint velocity tracking error is square-integrable may also be interpreted as ``fast enough'' servoing in the engineering sense.

\section{Simulation Results}

In this section, we provide numerical simulation results to demonstrate the performance of the proposed adaptive schemes using six standard two-DOF (degree-of-freedom) planar robots moving in the horizontal X-Y plane. The interaction graph among the robots is shown in Fig. 1. %The physical parameters of the 2-DOF robot are listed in Table I, where $m_i$, $I_{C,i}$, $l_{C,i}$, and $l_i$ represent, respectively, the mass, the moment of inertia with respect to the center of mass (CM), the length of the CM with respect to the prior joint, and the length of link $i$, $i\in \left\{1,2,E\right\}$, where the tool is considered to be link $E$.
 The sampling period is chosen as 5 ms.

\subsection{Dynamic Controller}

We first consider the dynamic controller given in Sec. III. The entries of the weighted adjacency matrix $\mathcal W$ are chosen as $w_{ij}=0.5$ if $j\in{\mathcal N}_i$, and $w_{ij}=0$ otherwise. The communication delays among the robots, for simplicity, are set to be $T_{ij}=0.5$ s, $\forall j\in{\mathcal N}_i,\forall i=1,2,\dots,6$.
The controller parameters $K_i$, $\Gamma_i$, $\Lambda_i$, $\alpha$, $\beta$, and $\lambda$ are chosen as $K_i=30.0 I_2$, $\Gamma_i=10.0 I_3$, $\Lambda_i=10.0 I_2$, $\alpha=10.0$,  $\beta=10.0$, and $\lambda=25.0$, respectively, $i=1,2,\dots,6$. The initial dynamic parameter estimates are determined as $\hat \vartheta_i(0)=\left[0,0,0\right]^T$, $i=1,2,\dots,6$. The initial kinematic parameter estimates are determined as $\hat \theta_1(0)=\left[1.5,2.5\right]^T$, $\hat \theta_2(0)=\left[3.2,3.2\right]^T$, $\hat \theta_3(0)=\left[2.6,2.8\right]^T$, $\hat \theta_4(0)=\left[3.2,2.7\right]^T$, $\hat \theta_5(0)=\left[3.5,2.9\right]^T$, and $\hat \theta_6(0)=\left[1.3,2.8\right]^T$. The initial values of the observed quantities are set as $x_{o,i}(0)=x_i(0)+0.02$, $i=1,2,\dots,6$. Simulation results are shown in Fig. 2 and Fig. 3, and the task-space positions of the robots indeed converge to the same value $\left[2.0837,0.5645\right]^T$.

%\subsection{Tracking Problem}
%
%\subsection{Regulation Problem}

%\begin{table}[!t]
%\renewcommand{\arraystretch}{1.2}
%\renewcommand{\tabcolsep}{0.12cm}
%\center\caption{The physical parameters of the robots}
%\begin{tabular}{c c c c c}
% \hline $i$-th robot & $m_1, m_2(kg)$& $I_{C,1}, I_{C,2} (kg\cdot m^2) $& $l_1,l_2(m)$& $l_{C,1},l_{C,2}(m) $ \\
%\hline
%
%    1 & 1.2,    1.6    &0.3610, 0.4813    &1.90, 1.90    &0.95, 0.95\\ \hline
%    2 & 1.8,    1.5    &0.7260, 0.4050    &2.20, 1.80    &1.10, 0.90\\ \hline
%    3 & 2.1,    1.6    &0.6318, 0.7053    &1.90, 2.30    &0.95, 1.15\\ \hline
%    4 & 2.2,    1.5    &0.8873, 0.3200    &2.20, 1.60    &1.10, 0.80\\ \hline
%    5 & 1.1,    1.6    &0.4043, 0.4320    &2.10, 1.80    &1.05, 0.90\\ \hline
%    6 & 2.3,    1.7    &0.6919, 0.5114    &1.90, 2.10    &0.95, 1.05\\ \hline
%\end{tabular}
%\label{TabI}
%\end{table}

\begin{figure}
\centering
%%----start of first figure----
\begin{minipage}[t]{1.0\linewidth}
\centering
\includegraphics[width=2.3in]{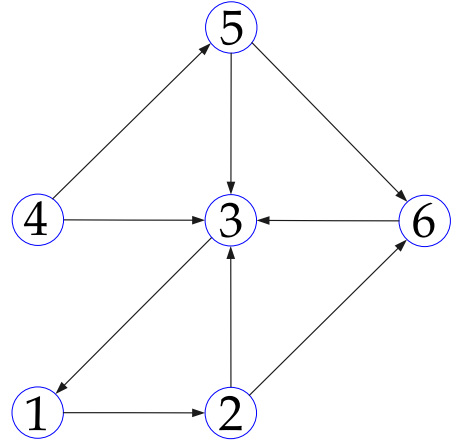}
\caption{Interaction graph among the robots.}\label{fig:side:a}
\end{minipage}%
\end{figure}

\begin{figure}
\centering
%%----start of first figure----
\begin{minipage}[t]{1.0\linewidth}
\centering
\includegraphics[width=2.8in]{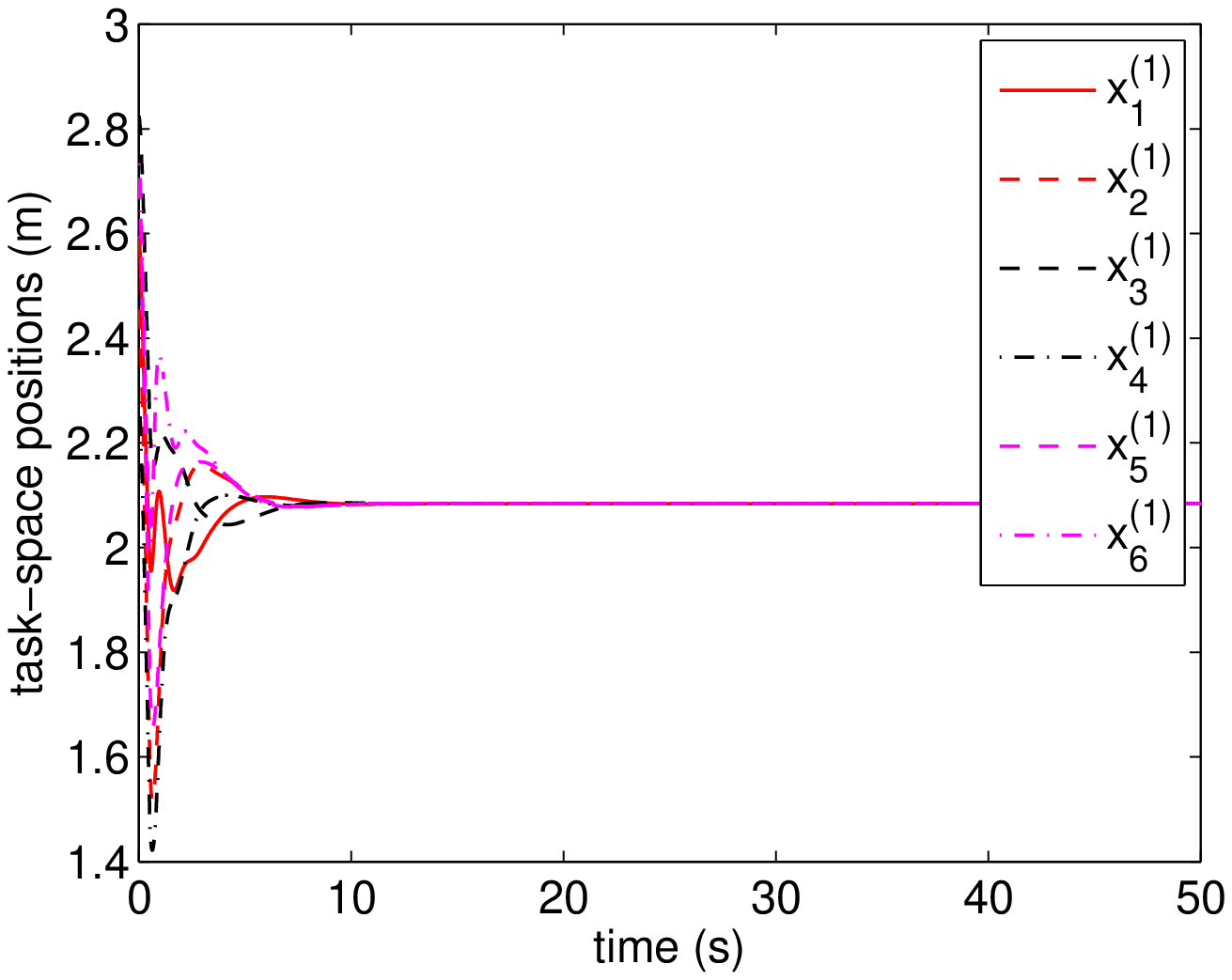}
\caption{Task-space positions of the robots (X-axis).}\label{fig:side:a}
\end{minipage}%
\end{figure}

\begin{figure}
\centering
%%----start of first figure----
\begin{minipage}[t]{1.0\linewidth}
\centering
\includegraphics[width=2.8in]{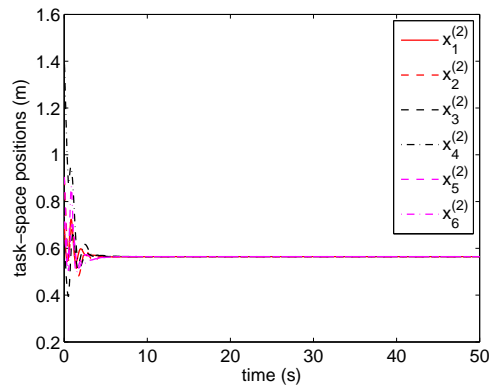}
\caption{Task-space positions of the robots  (Y-axis).}\label{fig:side:a}
\end{minipage}%
\end{figure}

\subsection{Dynamic Controller With an External Input/Stimuli}

{
Suppose that the 1st robot is subjected to an external input $\tau_{h,1}=J_1^T(q_1)[-15\dot x_1-30(x_1-x_h)]$ (exerted by a human operator) after $t=10$ s with $x_h=[2.6,0.9]^T$ denoting the desired position. Under the same context and with the same controller parameters, the task-space positions of the robotic systems are shown in Fig. 4 (X-axis) and their X-axis positions finally stay at 2.1111 rather than 2.6. To increase the manipulability of the system, we decrease $\alpha$ from $\alpha=10.0$ to $\alpha=0.05$. The corresponding simulation results are shown in Fig. 5 and the average of the X-axis positions of the robots finally stays at 2.5347, which is much closer to 2.6 in comparison with the case $\alpha=10.0$. In the extreme case $\alpha=0$ (i.e., the manipulability index is infinite), the responses of the task-space positions of the robots are shown in Fig. 6 and the final positions of the robots (the average of the X-axis positions is 2.5979) are very close to the the desired one $x_h^{(1)}=2.6$.}
\begin{figure}
\centering
%%----start of first figure----
\begin{minipage}[t]{1.0\linewidth}
\centering
\includegraphics[width=2.8in]{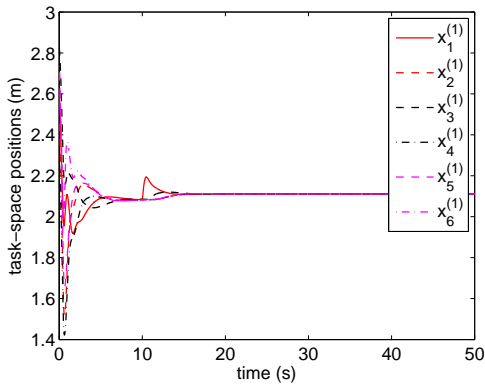}
\caption{Task-space positions of the robots (X-axis) under an external input and with $\alpha=10.0$.}\label{fig:side:a}
\end{minipage}%
\end{figure}

\begin{figure}
\centering
%%----start of first figure----
\begin{minipage}[t]{1.0\linewidth}
\centering
\includegraphics[width=2.8in]{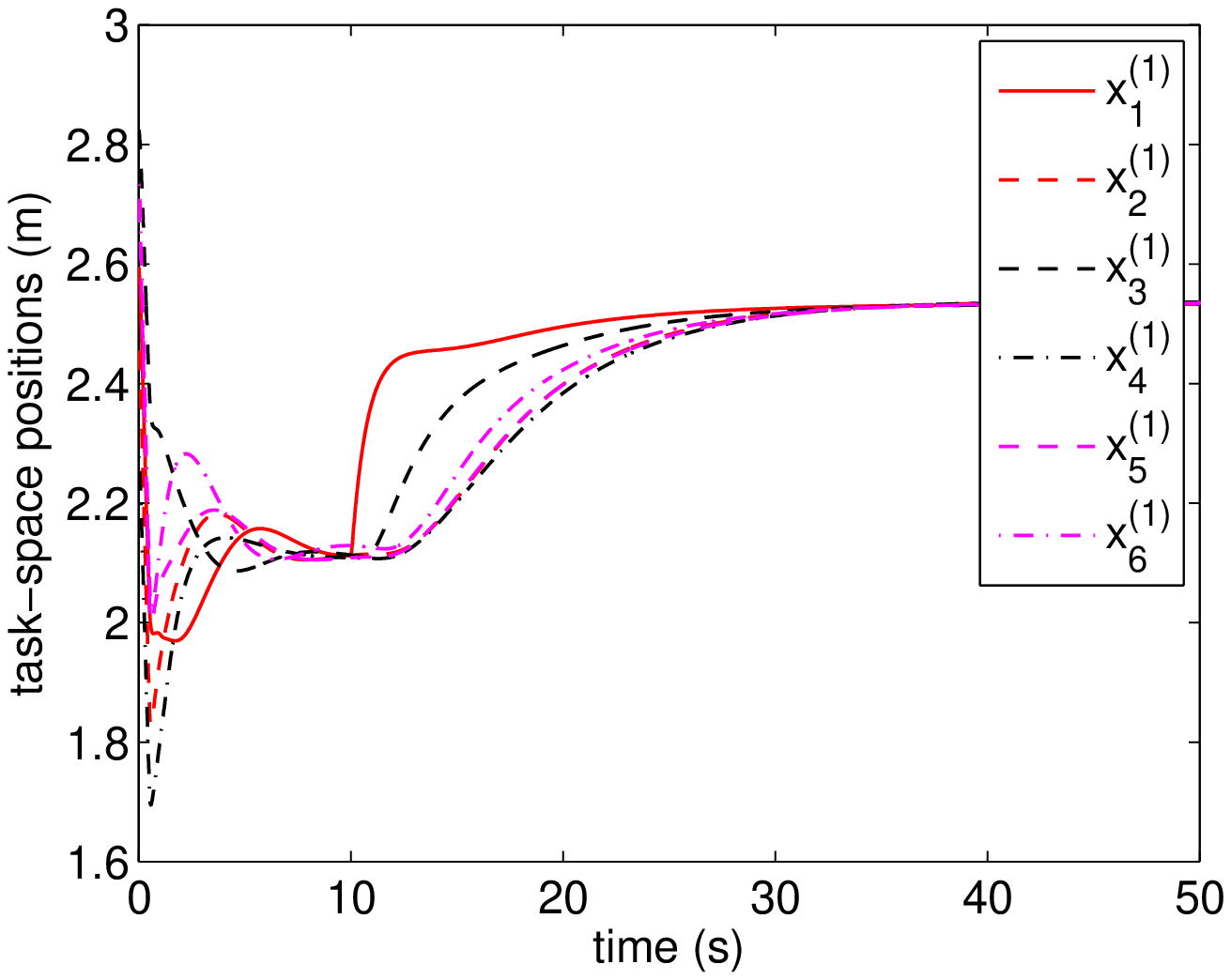}
\caption{Task-space positions of the robots (X-axis) under an external input and with $\alpha=0.05$.}\label{fig:side:a}
\end{minipage}%
\end{figure}

\begin{figure}
\centering
%%----start of first figure----
\begin{minipage}[t]{1.0\linewidth}
\centering
\includegraphics[width=2.8in]{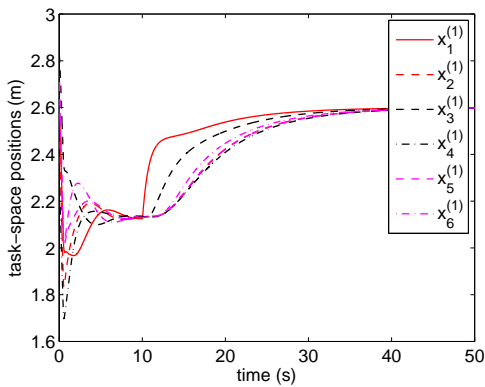}
\caption{Task-space positions of the robots (X-axis) under an external input and with $\alpha=0$.}\label{fig:side:a}
\end{minipage}%
\end{figure}

\subsection{Kinematic Controller and Discussion of Manipulability}

In this last scenario, we suppose that a PI (proportional-integral) velocity controller is embedded in the six robots (similar to most industrial/commercial robots) with their PI gains being set as $\bar K_P=60.0I_2$ and $\bar K_I=10.0I_2$. Under the proposed kinematic controller in Sec. IV with the parameters being chosen to be the same as those in Sec. V-A except that $\alpha=0$, the simulation results are shown in Fig. 7, and the X-axis task-space positions of the robots finally stay at 2.3762.
This difference from the case of using a dynamic controller is due to the integral action of the low-level PI velocity controller (in fact the existence of an integral action gives rise to the consequence that the gain of the mapping from the external force to the consensus equilibrium increment is bounded rather than infinite). To see this clearly, we set the integral gain $\bar K_I=0$, i.e., removing the integral action, and the simulation results are shown in Fig. 8 with the X-axis task-space positions of the robots finally staying at 2.59 (approximately) which is quite close to the desired value 2.6 (but with a relatively slower response in comparison with the dynamic controller---see Fig. 6). These simulation results partly demonstrate that the gain $\alpha$ of the kinematic controller and the integral gain $\bar K_I$ of the low-level PI controller simultaneously quantify the manipulability of the system.

In order to further investigate the robustness of the kinematic controller with respect to measurement noise, we perform a simulation with measurement noise being added to joint positions, joint velocities, and task-space positions. The measurement noise of the joint positions conforms to a normal distribution with its mean being zero and its standard deviation being 0.002 rad and that of the joint velocities conforms to a normal distribution with its mean being zero and its standard deviation being 0.005 $\text{ rad$\cdot\rm {\text s}^{-1}$}$. The measurement noise of the task-space positions conforms to a normal distribution with its mean being zero and its standard deviation being 0.01 m. Simulation results are shown in Fig. 9, which demonstrate that the proposed kinematic controller indeed has certain robustness with respect to the measurement noise.

 %One may need to be cautious about the choice of the PI gains, and in fact $K_P$ should be set to be high enough, to an extent that the joint servoing is fast as is required in Theorem 2.

\begin{figure}
\centering
%%----start of first figure----
\begin{minipage}[t]{1.0\linewidth}
\centering
\includegraphics[width=2.8in]{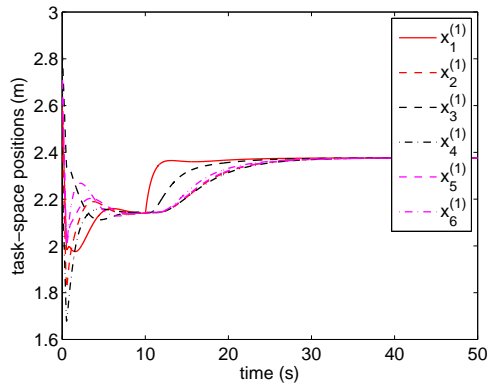}
\caption{Task-space positions of the robots (X-axis) under a kinematic controller and an external input and with $\alpha=0$.}\label{fig:side:a}
\end{minipage}%
\end{figure}

\begin{figure}
\centering
%%----start of first figure----
\begin{minipage}[t]{1.0\linewidth}
\centering
\includegraphics[width=2.8in]{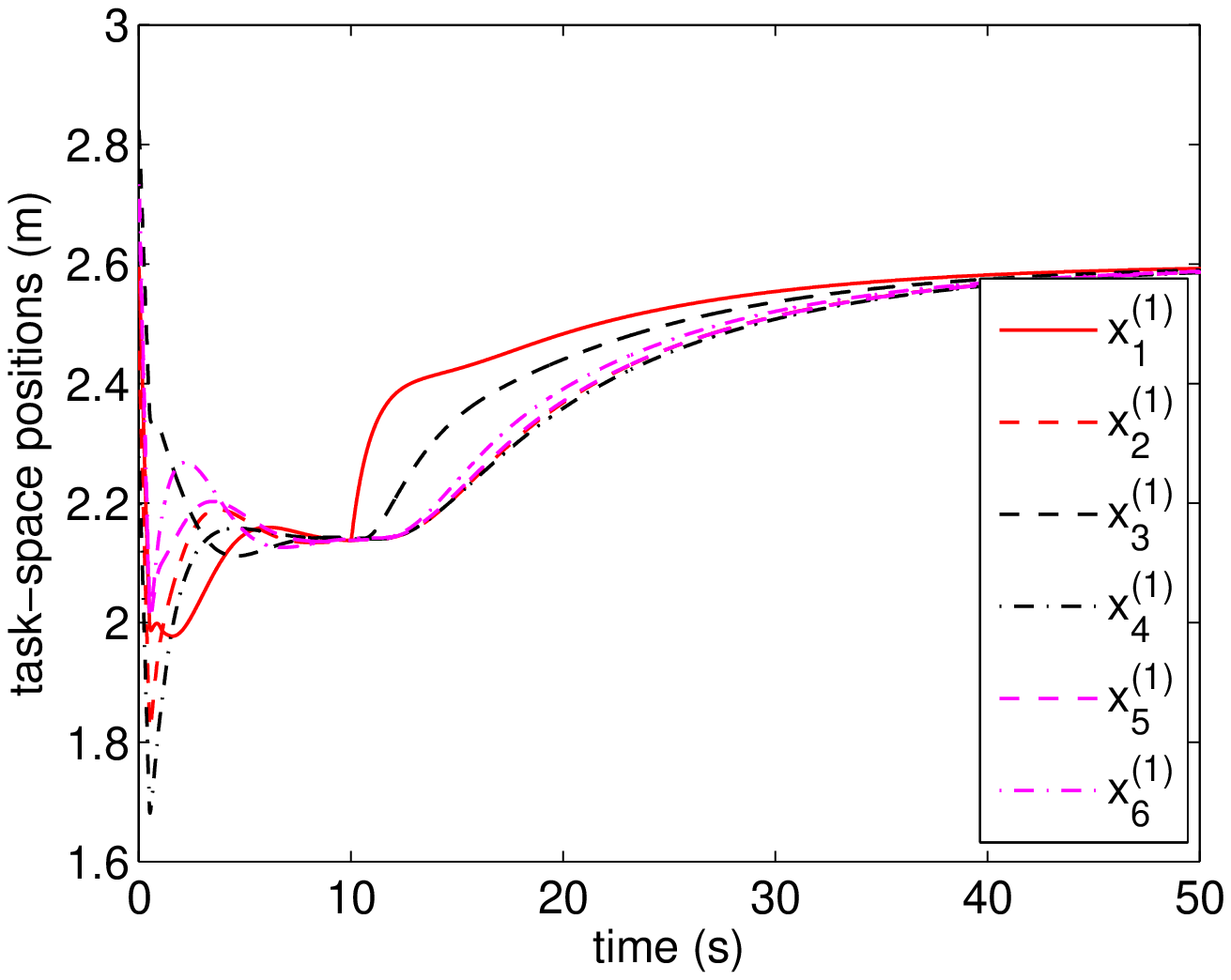}
\caption{Task-space positions of the robots (X-axis) under a kinematic controller and an external input and with $\alpha=0$ and $\bar K_I=0$.}\label{fig:side:a}
\end{minipage}%
\end{figure}

\begin{figure}
\centering
%%----start of first figure----
\begin{minipage}[t]{1.0\linewidth}
\centering
\includegraphics[width=2.8in]{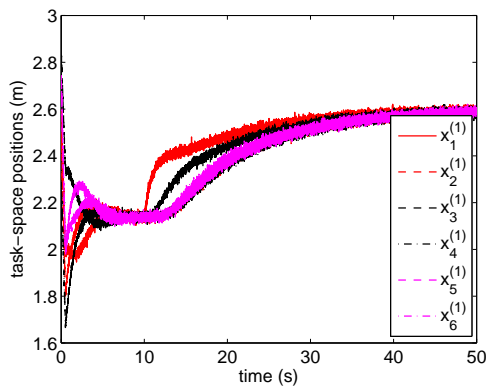}
\caption{Task-space positions of the robots (X-axis) under a kinematic controller and an external input and with $\alpha=0$ and $\bar K_I=0$ (measurement noise is considered).}\label{fig:side:a}
\end{minipage}%
\end{figure}

\section{Conclusion}

In this paper, we have studied the task-space consensus problem for multiple robotic systems with kinematic and dynamic uncertainties. We propose an observer-based adaptive consensus scheme to achieve the manipulable consensus objective without relying on the task-space velocity measurement. The main new features of our work are that 1) the observer relies on the joint reference velocity rather than the joint velocity, 2) the kinematic parameter adaptation law uses a distributed adaptive kinematic regressor matrix and is driven by both the observation and consensus errors, 3) the separation of the kinematic and dynamic subsystems is achieved without involving task-space velocity measurement, and 4) {the concept of manipulability is formalized and the manipulable consensus of the robotic systems is ensured}. One general conclusive result obtained is that the networked systems with strong manipulability typically contain a mapping with an infinite gain from the external force/torque to the consensus equilibrium increment, and that the strong/infinite manipulability provides the possibility of adjusting the consensus equilibrium in an arbitrary range by finite control efforts of the human operator (implying the less fatigue of the human operator). The performance of the proposed adaptive schemes is shown by numerical simulation results in both the case of open torque design interface and that of closed controller architecture. %Future research will be devoted to addressing the  the consideration of the much challenging time-varying communication delays.

% or


%\appendices

%\section{}

\section*{Acknowledgment}

The authors would like to thank Prof. Chien Chern Cheah for the helpful suggestions on improving the quality of the paper, Dr. Yang Zhou for the valuable discussions concerning the manipulability of commercial robots with an inner PI velocity controller, and Dr. Huang Huang for the helpful comments on improving the presentation concerning the manipulability analysis based on the input-output stability.

%\ifCLASSOPTIONcaptionsoff
%  \newpage
%\fi



\bibliographystyle{IEEEtran}
\bibliography{..//Reference_list_Wang}

%\begin{thebibliography}{1}
%
%
%\end{thebibliography}

%\begin{biography}[{\includegraphics[width=1in,height=1.25in,clip,keepaspectratio]{mshell}}]{Michael Shell}

%\end{IEEEbiography}


%\newpage



%\vfill


\end{document}